# Complex Philosophy

Carlos Gershenson
School of Cognitive and Computer Sciences
University of Sussex
C.Gershenson@sussex.ac.uk
http://www.cogs.susx.ac.uk/users/carlos

## Abstract

*We present several philosophical ideas emerging from the studies of complex systems. We make a brief introduction to the basic concepts of complex systems, for then defining "abstraction levels". These are useful for representing regularities in nature. We define absolute being (observer independent, infinite) and relative being (observer dependent, finite), and notice the differences between them. We draw issues on relative causality and absolute causality among abstraction levels. We also make reflections on determinism. We reject the search for any absolute truth (because of their infinity), and promote the idea that all comprehensible truths are relative, since they were created in finite contexts. This leads us to suggest to search the less-incompleteness of ideas and contexts instead of their truths.*

## 1. Introduction

As science, knowledge, and ideas evolve and are increased and refined, the branches of philosophy in charge of describing them should also be increased and refined. In this work we try to expand some ideas as a response to the recent approach from several sciences to complex systems. Because of their novelty, some of these ideas might require further refinement and may seem unfinished[1], but we need to start with something. Only with their propagation and feedback from critics they might be improved.

We make a brief introduction to complex systems, for then defining *abstraction levels*. Abstraction levels represent simplicities and regularities in nature. We make an ontological distinction of absolute being and relative being, and then discuss issues on causality, metaphysics, and determinism. These are in general philosophical ideas, but they should be interesting for researchers of complex systems, since they have implications in the way we observe complex systems.

## 2. Complex Systems

Since complex systems can be found almost everywhere and in a wide variety of contexts, it is very difficult to abstract them into a well-defined, crisp-bounded concept. But a loose-defined, fuzzy-bounded concept is good enough for our purposes. More specific definitions can be made in specific contexts. So, a complex system consists of elements, which interact with each other, with global properties of the system which are not found on the elements. These properties are said to **emerge** from the interactions of the elements. The complexity of the system is proportional to the number of elements it has, to the number of their interactions, and to the complexities of the elements and the complexities of their interactions.

Let us now see some examples of complex systems.

- A **cell** is formed by proteins and molecules, which are not considered to be alive. But the elements of the cell are *organized* in such a way that as external observers, we judge that the cell is alive. Life *emerges* from the interactions among different proteins and molecules.
- A **brain** consists of billions of neurons. A single neuron is not capable of controlling the body of an animal, while neurons organized in a nervous system are capable of providing adaptation to animals, and in some cases intelligence and consciousness. All these *emerge* from the neurons' interactions.
- A **society** presents many properties that its members cannot have by themselves, such as collective behaviours, beliefs, and misbeliefs, that may *emerge* from simple interactions among the members (Gershenson, 2001).
- **Cellular automata**, such as the "Game of Life" (Conway, 1982; Gershenson, 1997; Wuensche, 1998), consist of matrixes where each element has a state or value. This state is modified through time taking into account the states of the neighbour elements. Very simple rules for modifying states yield to emergent complex global behaviour in the system.

---

[1] Is there such a thing as a "finished idea"?

A complex system may consist of only two elements (which in turn might be also complex systems). An example could be a symbiotic relationship between two animals. Each animal would not survive as it does if it would not be because of the relationship with the other, so we can say that their survival emerges from their interactions.

One of the main reasons for studying complex systems is that this approach allows us to understand the behaviour of the system by understanding the behaviours and interactions of the elements. Following Newtonian determinism, this lead people to believe that if we could understand the "simple basic elements" of the world (similar to the Greek concept of atom), we could be able to understand all the world. But physicists in search of these "simple basic elements" have found more and more complexity in subatomic particles. Well, there is no reason for why shouldn't we be able to divide anything, no matter how small it is. So, in theory, we could say that we will never stop finding smaller elements of our world, never finding the "real" atoms[2].

But if we can find simple phenomena in different contexts, ¿where does this simplicity comes from? Similar to complexity, it also emerges. So we can speak about **emergent simplicity** and **emergent complexity** (Bar-Yam, 1997). In a system, when the number of elements and interactions is increased, the emergent complexity is also increased, but not *ad infinitum*. Many complex systems are characteristic because they present **self-organization**. These systems are called **complex adaptive systems** (CAS). Self-organization is given when regularities begin to occur in the system. These regularities give rise to emergent simplicity. How these regularities occur is a very interesting question addressed by researchers, but there is more than one answer in dependence of the system. Some regularities arise by bounds or limits of the system or of its elements, others emerge from local rules that yield to a uniform behaviour of the elements, and others we still do not have a clue.

## 3. Abstraction levels

We do not know exactly how concepts are created in our minds[3], but we believe that they arise from regularities in perception. Therefore, "simple" systems will have a well-defined concept representing them because of the regularities in the system. On the other hand, the more complex a system is, the harder it will be to understand and to have clear concepts of it.

In our world, we can perceive simplicity at different levels (*e.g.* atoms, proteins, individuals, planets, galaxies). We call these **abstraction levels**. Their regularities and relative simplicity allow us to have clear concepts of them, even when they might be composed of many complex systems, because their global behaviour is "simple". Elements represented in abstraction levels interact giving emergent complexity, until new regularities arise, and we can distinguish another abstraction level. Figure 1 shows some abstraction levels that can be identified. From an abstraction level a complexity level arises by emergent complexity, and then another abstraction level arises from the complexity level by emergent simplicity. **Complexity levels** represent the complexity perceived in objects and phenomena, but they are not well defined concepts. We draw two complexity levels below subatomic particles because we suppose that they are divisible, and thus be a product of emergent simplicity, but there is no first abstraction level. We believe that the universe (in the etymological sense, this is, *everything*) is infinite (and if not, it behaves as if it would be (we have not found limits)). Therefore, it can be considered as an abstraction level and as a complexity level at the same time. We can have a concept of it, but not because of its regularities, but because of generalization. And the behaviour of the universe as a system can be as complex *and* as simple as we want to define it, because we cannot perceive such behaviour. We call "big bangs" to phenomena similar to the one it seems most of the matter we can perceive came from. But since we believe the universe has no limits in space or time (what would be on the other side, then?), we believe that there should be an infinite number of big bangs in different stages. Most people call our big bang universe, but as we said, we understand for universe *everything*.

---

[2]"Atom" comes from the Greek "τομή" (division) and with the negative prefix "a" literally means "indivisible".

[3]This also depends on our concept of "concept". We define a concept as a generalization of perception(s) or other concept(s).

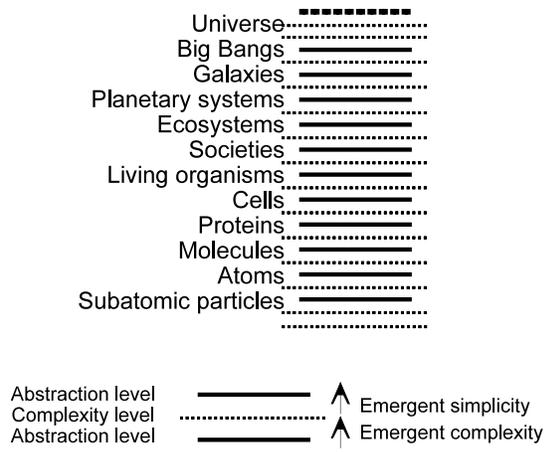

Figure 1. Some abstraction levels

Let us remember our loose and recursive definition of complexity: "the complexity of the system is proportional to the number of elements it has, to the number of their interactions, and to the complexities of the elements and the complexities of their interactions". Just like this, the definition is far from being practical. Well, since we cannot find a "first" abstraction level, the recursion in the definition has no end, so we cannot speak of an **absolute complexity**. But we can take any abstraction level as a point of reference that fits our needs, and then we can have a finite recursion, but we would speak of a **relative complexity**. Emergent simplicity does not affect the definition of complexity, but it should affect the reference point. For example, we could say that the behaviour of a planet orbiting around a single star is less complex than protein interactions in the cell. But this is because the planet is simple at an abstraction level, and this is its reference point. The proteins are far from their closest lower abstraction level because of their number and interactions. But if we want to study a planet in terms of proteins, it would be much more complex than a cell.

It has not much sense in studying, following the previous example, a planet in terms of proteins, or quoting Herbert Simon, a sheep in terms of quarks. This is not because a quark cannot affect the behaviour of the sheep, but because the effect of the quark can be perceived in *every* abstraction level until reaching the sheep. Causality cannot jump abstraction levels. But can we speak about causality among abstraction levels? Yes, but very carefully. First we need to make a small distinction.

## 4. Ontology

We can define two types of **being**: absolute and relative. Let us call the absolute being **a-being** and the relative **re-being**. The a-being is the being which is independent from the observer, and is for and in all the universe. Therefore, it is infinite and uncomprehensible, although we can approximate it *as much as we want to*. The re-being is the being which is for ourselves, and it is different for each individual, and therefore dependent from the observer. It is relative because it depends on the *context* where each individual is, and this context is different for all individuals, and even the context of an individual is changing constantly, with his or her representations of what *re-is*. The re-being depends on experience, reason, and beliefs, which in turn depend on each other.

*Objects do not depend on the representation we have of them*. The re-being depends on the a-being, but the a-being... well, just *a-is*, independently from any observer. A table may re-be nice and decorative for one person, and the same table may re-be small but practical for another person, and re-be ugly, tasteless and fragile for a third person. But it a-is the same and one table, *independently* of what it re-is for anyone. This does not imply that the a-being cannot change in time nor be dynamic. If the table burns it will a-be a different thing that the one it a-was, also independently of what the burnt table re-is for anyone. Figure 2 shows a graphical representation of the ideas presented above: the a-being contains different re-beings, which in turn might contain others, intersect, or be excluded. No re-being can contain or be equal to the a-being, because this last one is infinite, while all re-beings are finite. The larger the re-being is, the less incomplete it is. The re-beings are dynamic, and we *believe* that the a-being is also dynamic.

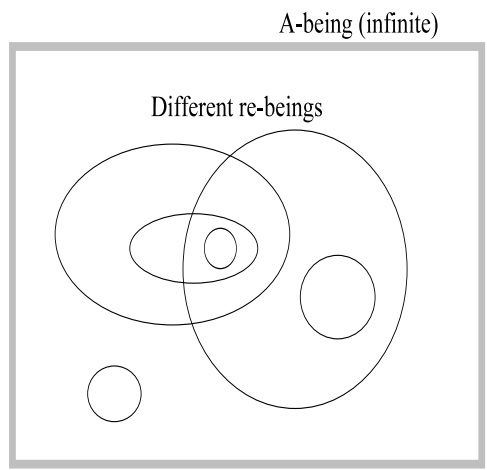

Figure 2. A-being and re-beings.

Another way of representing these ideas would be with the following example: we have a ball, which *a-is* 50% white and 50% black. But we can only see it from one perspective. So, for some of us the ball will *re-be* completely white, for others it will *re-be* completely black, or 70% black and 30% white, etc. This is because each one of us has a different perspective (context). And our contexts do not affect the colour of the ball. We can only get an idea of the complete colour of the ball by taking into account as much perspectives (re-beings) as we can. We cannot make an average, since most of us could be watching it from the same perspective (for example, we might conclude that it is 99% white). (Note that a usual ball is not infinite as the a-being is). Figure 3 shows three perspectives of this supposed ball.

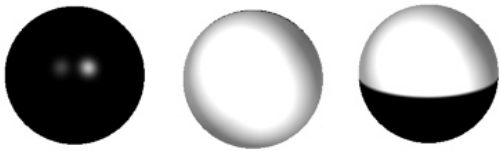

Figure 3. Which colour the ball a-is?

The **being** would be the conjunction of a-being and re-being. Confusing? Well, it would explain some centuries of debates... For example, the old proverb: "if a tree falls in the forest and no one is there to hear it fall, has it really fallen?". Well, the tree a-has fallen, but noone could be able to represent it, so it re-has not fallen for anyone. So the answer if we do not make the previous distinction would be simply yes *and* no. Another example can be found in the debate between empiricism and rationalism. When empiricists spoke about "being", they referred to something closer to the a-being, because things "were" (a-were) independently from the observer, and we needed of experience to perceive the "being" (a-being) of things. And when rationalists spoke about "being", they seem to have referred to something closer to the re-being: "I think, therefore I "am" (re-am)". No wonder why there was a strong debate if they were speaking about different things trying to use the same concept.

## 5. Causality

Returning to our discussion about abstraction levels: can there be causality among them? There *re-is*, but there *a-is* not. Let us elaborate this idea. There can re-be causality among levels because *we* define the abstraction levels. We define the concepts that are used for identifying causality. For example, from one point of view the behaviour of a society depends on the behaviour of the individuals, but from another point of view the behaviour of the individuals depends on the behaviour of the society. We can speak about this if we have a reference point, individuals in the first case and societies in the second. But the a-being has no reference point. There a-is no causality between individuals and societies, because they *a-are the same thing*. Individuals and societies *re-are* concepts that re-were abstracted because of their regularities, but independently from us, a society and the members that compose it a-are the same thing[4]. Generalizing, if we take a concept from a certain abstraction level, the elements from lower abstraction levels that compose it a-are the same thing. For example, a planet *a-is* all the atoms of the planet interacting between them, producing several emergent complexities and simplicities, allowing us to perceive several abstraction levels. So, among abstraction levels we can speak about **relative causality**, but not about **absolute causality**[5].

Can we speak about matter, then? Everything seems to *a-be* only emergent properties of smaller complex systems, but there is not a "basic essential" class of elements. *Everything **a-is** an infinitude of nothings*[6].

*There **a-is not** an essence in the universe. Everything **a-is** the essence. Everything **a-is** based on everything, everything **a-is** related*. There can *re-be* an much as essences as we want, because we can take as a reference point any concept we feel fit for being an essence. Or, we could speak about a *circular-relative causality*, because if we set as an essence a reference point, we will return to that reference point if our context is complete enough (*e.g.* individuals cause states of societies, and societies cause states of individuals; atoms cause states of molecules, molecules cause states of proteins, proteins cause states of cells, cells cause states of proteins, proteins cause states of molecules, molecules cause states of atoms). If we expand our contexts enough, we will approach what we just said: *everything is based on everything*, because we will see that any abstraction level has a causality (direct or indirect) on any other abstraction level. Because everything *a-is* the same *thing*. Of course, for studying and understanding

---

[4] We are not saying that independent individuals have the same capabilities than their societies, but that all the individuals, interacting, *a-are* the society.

[5] Remember we are speaking about causality among abstraction levels, not about causality in time.

[6] I declare myself a monist, but not a materialist (matter for me is also an emergent property).

our world we need to make use of our abstractions, and set borders to our contexts.

And if we want to speak about causality in time, we can see that it becomes very difficult to observe as the complexity of a system increases. Mainly because of the interactions among the elements of a complex system determine the behaviour of other elements. The more there are interactions, the more causes the behaviour of an element will have. When a system is complex enough, the causes become untraceable. If we try to find "the" cause of something, very probably we will find ourselves trapped in a "chicken-egg" "problem", because if the interactions of a complex system are very high. For example, if A is caused by B, C, D, E, F, G, and H, it is very probable that any of these will be partly caused by A. Mathematical models, such as random boolean networks (Kauffman, 1993; Gershenson, in preparation), show very clearly how the complexity (and the tractability) of a system increases proportionally to the number of interactions among elements. Therefore, in cases of extreme complexity, we cannot study causality *of the elements* in time, since *everything depends on everything*. We can only study the system as a whole, and try to understand the properties which arise by emergent simplicity. Only at a system level we might be lucky enough to find if a state of the system causes another state of the system.

## 6. Metaphysics

Metaphysics can be seen as the axioms of philosophy and thought. That is to say, metaphysics are the ideas we **believe** in and base our reasoning upon. Our reason cannot *prove* our beliefs in the same way that theorems derived from axioms cannot prove the axioms (Gödel, 1931; Turing, 1936).

For thinking and reasoning, we need basic ideas in order to begin thinking. These basic ideas would be our metaphysics. We can then build new ideas over our metaphysics, but this does not mean that our metaphysics cannot change, even by the ideas based on them.

But if in the universe there a-is no essence, <and|or>[7] everything a-is the essence, how can our metaphysics, and all the ideas based upon them be valid? They a-are not. They cannot a-be, because they are finite, and our universe is not[8]. The question is to see if they re-are valid. And the answer will be always yes, because the ideas were created according to a specific context, and they fit in that context. People do not make non-valid ideas by their own will. *Every idea is valid in the context it is created*. It is when we take an idea out of its context that it might or not might be valid. Since contexts are dynamic, when we make an error it is because we expanded our context by perceiving something which we were not aware of while we make an error. Only after expanding our context we can realize that we had an error, since it becomes non consistent with the expansion of our context which makes us realize of our error.

So all ideas have the same degree of validness? Of course not. We can say that an idea is **less incomplete** as it is valid in more contexts. The more contexts an idea is valid in, and the wider these contexts are, will make the validness of the idea higher. *An idea will never be complete*, but we can make our ideas *as less incomplete as we want to*. We can see that the search for truth is obsolete if it is not *relative* to a reference point, because an absolute truth a-is infinite.

With this we can explain why there have been different explanations for the same thing. People look at things from different contexts. Their context is finite and *re-is* their personal essence. Therefore, they can explain things in different ways. Which way is the "true" one? We cannot say if it is not related to a reference point. If we have two ideas from different contexts, explaining the same thing, to see which one is "better", we first need to see "better for what". An idea will be better for something if that something is closer to its context than to the context of the other idea. An easy example: which one is "better": neoliberalism or socialism? Well, we need to specify better for whom. Each one is better for people who obtain benefits from one or other. Neoliberalism re-is good for businessmen, socialism re-is good for workers. In this relativistic context, we can say that Prothagoras and Methrodorus were right: "Man *re-is* the measure of all things", "All things *re-are* what people think of them". If someone asks: which one a-is better, neoliberalism or socialism? I am afraid we could not answer without falling into imprudence. It seems that a system which would benefit both businessmen and workers would re-be better than both socialism and neoliberalism. And such a system would be less incomplete, because it would be valid in the contexts of the workers and of the businessmen.

---

[7]We use the notation <A|B> when we mean "A and B at the same time".

[8]Let us note that we are not saying that metaphysics is not useful or needed. On the contrary. We are just noticing its limits.

Different ideologies arise from different contexts. But all contexts are incomplete. There cannot be a completely valid ideology. The ideas exposed here are also dependent of their context. Once our culture evolves, and contexts are enhanced, their incompleteness will make them obsolete. Any context, no matter how less-incomplete it is, it will never be complete, so it will not stop being **relative**. It is fortunate, otherwise we would kill Philosophy.

So, if all ideas are relative to their context, and none can be absolutely valid, how can we make science? Well, it seems that science does not intend to *find* absolute truths, but to *approximate* them. Many ideas are valid in all our contexts. We still cannot say that they a-are valid, but we have a very high certainty of them. Anyway, if they re-are valid in all our contexts, and they re-are not valid in others we do not know about, we can be indifferent about it.

## 7. Determinism

Let us return to our statement from § 5, "everything is an infinitude of nothings". Is there determinism[9] then? Well, we cannot say if there a-is or a-is not, but we can see that there can re-be. Taking the popular Heisenberg example from quantum mechanics: if we cannot know the position *and* the velocity of a particle without changing it, the behaviour of the particle **re-is** *uncertain*. If the behaviour of all particles in the universe re-is uncertain for us, can we speak of determinism? It can re-be, because of emergent simplicity. The same regularities that allow us to create abstraction levels make the phenomena we call "simple" deterministic. Subatomic particles or quarks may not a-be deterministic *per se*, but their uncertainty does not affect us. We do not know if they a-are deterministic or not. Since they produce emergent simplicity, we can perceive determinism in systems emerging from their interactions. Also it seems that non-deterministic behaviour is much harder to abstract. We can say that there is a **relative determinism** where we have abstraction levels, which tends to re-be non-deterministic as it reaches a complexity level. In this case, by non-deterministic we do not mean random, but incomputable in a practical time[10]. We cannot determine the behaviour of the system not because we cannot know

"how it works", but because its complexity exceeds our computing or perceptual capacities.

Are the emergent properties of a complex system deterministic? We could say that they are if the elements of the system from which the properties emerge are deterministic. But since there a-is no "first" level of abstraction, we cannot say if things a-are deterministic or not. But because of emergent simplicity there are regularities in abstraction levels that re-are deterministic to our eyes. So we can say that phenomena re-are deterministic if our models are congruent with our perceptions "completely". This means that models simulate all the properties we perceive, which does not imply that models a-are congruent with things.

We could say that one of the objectives of science is to close the breaches between abstraction levels. This does not mean that we will understand an abstraction level completely, precisely because abstractions levels are incomplete, but again, we can make them as less incomplete as we want to. It does not matter how much we increase our computing capacities. A computer, a subset of the universe, cannot contain more information than the universe itself[11].

## 8. Conclusions

If science is enlarging our contexts, we need to update our ideas in order to make them valid for our ever expanding contexts. The present work is an attempt for achieving this in the case of complex systems.

The distinction made between *a-being* and *re-being* clarifies the indistinct and even more ambiguous use of "being". Apart from the ideas exposed, we can conclude from this distinction things already noticed before, for example that we cannot speak of an absolute good or evil (Schopenhauer, Nietzsche), but only of good and evil *relative* to a reference point. We have also developed ethics and aesthetics making this distinction of being.

But how to deal with the contradiction of having two types of being? Logic cannot accept contradictions. Well, instead of trying to change situations according to our logic, let us change logic according to our situations. "Logic is just a tool for reasoning, and does not determine what things *are*" (Schopenhauer, 1819). Paraconsistent logics (Priest and Tanaka, 1996) can deal with contradictions. An example of them is multidimensional logic (Gershenson 1998; 1999), where instead of having a truth value for a proposition, we have a truth vector whose elements may be contradictory. The result is that we can handle contradictions by adding dimensions to contain them.

---

[9] We shouldn't confuse determinism with predictability. Some chaotic systems are deterministic but they are not predictable.

[10] Lyapunov exponents too high, making a system too sensitive to initial conditions to be predictabe.

[11] Or could it? Only if the universe would be *self-affine* (Mandelbrot, 1998), but it seems it is not...

We stated that the ideas exposed here will not be valid as our contexts evolve. If they become invalid, then our prediction will be true. If they not, the ideas will be valid. This is a paradox, but it can be comprehended with paraconsistent logics. And it seems that if we want to be less incomplete, paradoxes cannot be avoided anymore, but they need to be comprehended[12].

Another important conclusion is that since all our ideas are relative, we should not search for the truth of our ideas, but for their less-incompleteness. All ideas are valid in the context they were created in, and no idea is completely valid, since our contexts are finite and our world is not.

We can conclude by saying that since all ideas are valid in the context they are created, we should tolerate ideas generated in contexts different than ours. Instead of denying something we do not comprehend, we should try to expand our contexts to include as much ideas as possible. The greater our contexts are, the less incomplete and more correspondent with reality they will be.

## 9. Acknowledgements

I would like to thank Nadia Gershenson, Antonio Gershenson, Gottfried Mayer-Kreiss, Jaime Besprosvany, Leon Lederman, and the members of the LOINAS group at Sussex University for their valuable comments.

This work was supported in part by the Consejo Nacional de Ciencia y Tecnologia (CONACYT) of Mexico and by the School of Cognitive and Computer Sciences of the University of Sussex.

---

[12]Using multidimensional logic operators (Gershenson 1998; 1999), we can see that paradoxes (which are true *and* false) are more "stable" than something which is only "true" or "false", because their negation is their equivalent (negation and equivalence are multidimensional logic operators). Multidimensional logic resources can be found at http://132.248.11.4/~carlos/mdl